# Phase statistics of light wave reflected from one-dimensional optical disordered media and its effects on light transport properties

Prabhakar Pradhan

*Department of Physics, University of Memphis, TN 38152*

Light wave reflection from optical disordered media is always associate with its phase, and the phase statistics influence the reflection statistics. We report a detailed numerical study of the statistics of the reflection coefficient $|R(L)|^2$ and its associated phase $\theta$ for plane electromagnetic waves reflected from one dimensional (1D) Gaussian white-noise optical disordered media, ranging from weak to strong disordered regimes. We solve numerically the full Fokker-Planck (FP) equation for the joint probability distribution in the $|R(L)|^2 - \theta$ space for different lengths of the sample with different "disorder strengths." The statistical optical transport properties of 1D optical disordered media are calculated using the full FP equation numerically. This constitutes a complete solution for the reflection phase statistics and its effects on light transport properties in a 1D Gaussian white-noise disordered optical potentials. Our results show the regime of the validation of the random phase approximations (RPA) or uniform phase distribution, within the Born approximation, as well as the contribution of the phase statistics to the different reflection averages for strong disorder regimes. Results of the previous work reported in the literature relative to the present work also been reviewed and discussed.

## 1. Introduction

Wave propagation through 1D random media have been studied for decades for both optical and electrical systems. Typical physical examples of these problems are light transport in random dielectric media, electron transport in disordered conductors, *etc.* All previous studies agree that the transmission through a 1D disordered medium decays with the length of the sample. For electronic systems, the conductance decays exponentially with the length of the sample and the resistance increases exponentially with the sample length. The main characteristic length scale in the problem is the localization length or transport mean free path. Results of different phases of research in this field have been reviewed recently in several review articles [1–7]. Disordered 1D systems are quite generic as regards to their transport properties — disordered 1D conductors are all alike, but every ordered 1D conductor is ordered in its own way! Thus, Mott et. al. [1] first showed that all the eigenstates of 1D disordered quantum systems are exponentially localized. There is no truly 1D disordered metal. For electronic systems quantum interference effects are important at low temperatures when de-coherence due to the inelastic process may be neglected. At low enough temperatures with static disorder, the transport properties of the electronic systems are highly fluctuating (i.e. sample specific) due to the different conductors with the same impurity concentration but with different microscopic arrangements of the impurities, can differ substantially in their transport properties. Several studies have shown that the root mean-squared-fluctuation is more than the average for length scales larger than the localization length, while for the length scales within the localization length (i.e. in the good metallic regime for quasi 1D and higher dimensional systems) the fluctuations are finite and universal, the so called universal conductance fluctuation (UCF) [7]. These fluctuations make the resistance and the conductance non-self-averaging quantities for electronic systems. The resistance at low temperatures when quantum effects are important is non-Ohmic, i.e., non additive in series. The non-additive nature arises because of the non-local effects of the quantum wave amplitudes associated with the electron in the conductor. Similar phenomena happen for the reflection coefficient in optical disordered media, which could be thought as an optical resistance. However, the reflection coefficient of an optical 1D system is bounded by the maximum reflection coefficient of the value 1. Due to this limitation, it will be shown later that the effect of non-self-averaging implied in the fact that the average and fluctuations for the reflection coefficient, both increase similar way as the electronic case, with fluctuation is higher than the average. Therefore, one has to take proper account of the phase for addition of the two optical samples. In this case, to calculate any meaningful quantity for a non-self-averaging quantity, like the reflection (for optical case), resistance and the conductance (for electronic case), one should know the full probability distribution of the same for every length of the sample. Several authors [8–14] have derived a Fokker-Planck (FP) equation for the full probability distribution of the reflection amplitude and its associated phase for the Gaussian white noise disordered optical or electronic media. Exact results for the average of different moments of the electronic conductance can be calculated analytically for the weak-disorder case using the Fokker-Planck equation only within the random phase approximation (RPA) (i.e., uniform phase distribution), for white noise disorder. However, for optical case, the FP equation is not closed in reflection coefficient within the RPA, and can be solved only with the limit of weak disorder and weak reflection. The approximation made in the weak disorder case is that the phase of the complex reflection amplitude relative to the incident wave, (or the relative phase of two non- Ohmically additive quantum resistances), are distributed uniformly, which is the random phase approximation (RPA). The analytical results for electronic cases show that the resistance and the conductance have log-normal distributions for large sample lengths. The characteristic length scale of the sample is called the localization length or transport mean free path [12-15]. Same things could be defined in optical case with a characteristic length scale of *optical localization length*. It is difficult to get an average quantity by direct analytical calculation for larger sample length which includes the complicated phase distribution, where the phase distribution depends on the strength of the disorder and length of the sample. However, the actual validity of the random phase approximation has not been studied in detail for varying disorder strength and sample length. Additionally, effect of the phase distributions on the reflection coefficients also not systematically addressed earlier. There are issues like actual contribution of the phase distribution to the different averaging processes that have not been systematically studied so far.

In this paper, we study systematically numerically, the joint probability distribution $P(r,\theta)$ of the phase ($\theta$) and the reflection coefficient ($r$) of the complex amplitude reflection coefficient ($R(L) = \sqrt{r}\, exp(i\theta)$) for one-dimensional optical disordered media with the Gaussian white noise disorder, for different lengths of the sample and different strengths of the disorder. The whole parameter space of the light reflection was explored. This is mainly the solution of the Fokker-Planck equation, where the probability density is varying in the $(r,\theta)$-space and evolving with the sample length $L$. Using the "invariant imbedding" technique, a non-linear Langevin equation can be derived for the $R(L)$. Then, using the stochastic Liouville equation for the probability evolution and the Novikov theorem to integrate out the stochastic aspect due to the random potential, one gets the Fokker-Planck (FP) equation in the $(r,\theta)$-space for varying sample length $L$ [10,12]. Integrating $\theta$ and $r$ parts separately from the joint probability distribution $P(r, \theta)$, we have calculated the marginal probability distributions $P(r)$ and $P(\theta)$ for $r$ and $\theta$, respectively, for a given length and disorder strength of the sample. We have outlined the range of validity of the random phase approximation, for the parameters on which the phase distribution depends, namely the sample length, localization length, and the wave vector of the incoming wave. We further analyzed the effect of the disorder on random phase approximation. We calculate the averages and the fluctuations of the reflection coefficients with different types of phase distributions associate with the disorder strength parameter of the optical samples. To the best of our knowledge, this is the first work where the joint probability distributions for the reflection coefficient and its phase for a 1D Gaussian white-noise disordered reflector are calculated for all length scales and for different disorder strengths, a complete visualization and quantification of the reflection probability for 1D disordered media. The way we have calculated the different quantities here involves essentially no approximations (within the numerical accuracy). We report here mainly the results of the light transport in random dielectric media (or, random layered media), but as discussed, the formalism apply equally well to the case of electronic transport in disorder conductors via Landauer formalism, and this shows that the phase statistics of the reflection coefficient and the phase associate with resistance/conductance are same. Finally, we also reviewed work of other on phase statistics comparison to the present numerical work.

## 2. Method of calculation

The Maxwell's equations can be transformed to the Helmholtz equation (wave equation), for both *E* and *H* fields. Consider the *E* part of the electromagnetic (EM) wave equation in scalar, polarization conserved, in a dielectric media with $\varepsilon = \varepsilon_0 + \varepsilon(x)$. Then *E* field in *1D* is given as:

$$\frac{\partial E^2}{\partial x^2} + \frac{\omega^2}{c^2}[\varepsilon_0 + \varepsilon(x)]E = 0 \quad . \tag{1}$$

Transforming the EM wave equation in E to the standard Helmholtz equation form, we obtain:

$$\frac{\partial u^2}{\partial x^2} + k^2[1+\eta(x)]u = 0, \qquad (2)$$

where $k = \varepsilon_0 \omega^2/c^2$ and $\eta(x) = \varepsilon(x)/\varepsilon_0$, where $\varepsilon_0$ is the constant dielectric background and $\varepsilon(x)$ is the spatially fluctuating part of the dielectric medium.

***The Langevin and The Fokker-Planck equations for the complex amplitude reflection coefficient:*** Consider a plane wave of wave vector $k$ incident from the right side of the disordered sample of length $L$ having the reflection amplitude $R(L)$. The non-linear Langevin equation for the reflection amplitude for the plane wave scattering problem can be derived by the invariant imbedding technique [12]. The Langevin equation for the complex amplitude reflection coefficient can be derived as:

$$\frac{dR}{dL} = 2ikR(L) + ik^2\eta(L)[1+R(L)]^2, \qquad (3)$$

with the initial condition $R(L)=0$ for $L=0$.

The main idea is to use the Langevin equation to get a FP equation for the reflection probability density. To get the Fokker-Planck equation from the non-linear Langevin equation [12], one has to get the probability density equation first and then the stochastic aspect due to the random potential has to be integrated out. The Langevin equation (i.e., Eq.3) can be solved analytically for the Gaussian white noise potential to get the FP equation. Detailed derivation have been reported several places, however, it is worth to repeat here some essential steps of the calculations for completeness of the numerical method. The non-linear Langevin equation for $R(L)$ in (Eq.(3) ) is basically a two coupled differential equations for the magnitude and the associated phase parts. Now, taking

$$R(L) = \sqrt{r(L)}\exp(i\theta(L)), \qquad (4)$$

and substituting Eq.(4) into Eq.(3), and equating the real and the imaginary parts on both sides of Eq.(3), one gets two coupled differential equations,

$$\frac{dr}{dL} = k\eta(L)r^{1/2}(1-r)\sin\theta, \qquad (5)$$

$$\frac{d\theta}{dL} = 2k + \frac{k}{2}\eta(L)\left[2 + \cos\theta(r^{1/2} + r^{-1/2})\right]. \qquad (6)$$

Now, according to the van-Kampen lemma [17], these two stochastic coupled differential equations will produce a flow of the density $Q(r, \theta)$ in the $(r, \theta)$-space and will obey the stochastic Liouville equation with increasing length of the sample, i.e., $Q(r, \theta)$ is the solution of the stochastic Liouville equation:

$$\frac{\partial Q(r,\theta,L)}{dL} = -\frac{\partial}{\partial r}(Q\frac{dr}{dL}) - \frac{\partial}{\partial r}(Q\frac{d\theta}{dL}), \qquad (7)$$

where $dr/dL$ and $d\theta/dL$ are given by Eqs. (5) and (6). Now, substituting the values of $dr/dL$ and $d\theta/dL$ in Eq.(7) one gets,

$$\frac{\partial Q(r,\theta,L)}{dL} = -k\sin(\theta)\frac{\partial}{\partial r}\left[r^{1/2}(1-r)\eta(L)Q\right] - k\frac{\partial}{\partial \theta}[\eta(L)Q]$$
$$-2k\frac{\partial Q}{\partial \theta} - \frac{k}{2}(r^{1/2}+r^{-1/2})\frac{\partial}{\partial \theta}[\cos\theta\eta(L)Q] \qquad (8)$$

To get the Fokker-Planck equation, Eq.(8) has to be averaged over the stochastic aspect, i.e., over all realizations of the random potential. For the case of a Gaussian white noise potential, Eq.(8) can be averaged out over the stochastic potential analytically using Novikov's [18] theorem. For the Gaussian white noise potential:

$$\langle \eta(L) \rangle = 0, \quad \langle \eta(L)\eta(L') \rangle = q\delta(L-L'). \tag{9}$$

Eq.(8) has terms like $\eta Q$ which are to be averaged out. For the Gaussian white noise disorder, the Novikov theorem states that:

$$\langle \eta(L)Q[\eta] \rangle = \frac{q}{2}\left\langle \frac{\delta Q[\eta]}{\delta \eta(L)} \right\rangle. \tag{10}$$

After averaging out the disorder aspect in Eq.(8) by using Eqs. (9) and (10), and writing $\langle Q(r,\theta) \rangle_\eta \equiv P(r,\theta)$, one gets the Fokker-Planck equation for $P(r,\theta)$:

$$\frac{\partial P(r,\theta,L)}{dl} = \left[\sin(\theta)\frac{\partial}{\partial r}r^{1/2}(1-r) + \frac{\partial}{\partial \theta} + \frac{1}{2}(r^{1/2}+r^{-1/2})\frac{\partial}{\partial \theta}\cos\theta\right]^2 P(r,\theta) \tag{11}$$
$$- 2k\xi\frac{\partial P(r,\theta)}{\partial \theta}.$$

Where $l \equiv L/\xi$, and $\xi = (qk^2/2)^{-1}$ is the localization length. The Fokker-Planck equation, Eq.(11), has all the information of the probability distribution of the reflection coefficient, $r$ and the associated phase ($\theta$) for different length scales of the sample and with varying disorder strengths. We will study systematically Eq.(11) numerically, which we are going to discuss in detail in later Sections.

### 3. Parameters of the problem

The Fokker-Planck equation (Eq.(11)) has three parameters to describe the problem fully: (i) the length of the sample $L$, (ii) the localization length $\xi$, and (iii) the incident wave vector $k$. In the re-arranged form of the Eq.(11), as it is written, it has effectively two parameters: $l = L/\xi$ and $C = 2k\xi$. Here $l$ is a number which gives the length of the sample in units of the localization length and $C$ is a number which fixes inverse of the disorder strength in terms of the wave vector of the incident wave and the localization length. Larger value of $C$ implies that the localization length is large, or the incoming electron energy is higher, or both, that is, the weak-disordered regime. Conversely, when $C$ is small it means $\xi$ is small, or the incoming wave energy is small or both, that is, the strong disorder regime. Intermediate/medium disorder regime is between the strong and weak disorder regimes.

### 4. Analytical solutions for *r* within the Random Phase Approximation (RPA), with weak disorder case

In the random phase approximation (RPA), which is valid for weak disorder and large incident optical/electrical field energies, one can write $P(r, \theta) = (1/2\pi)P(r)$ i.e., $P(r, \theta)$ factorizes, and $\theta$ is uniformly distributed over $2\pi$. Considering $\partial P/\partial \theta = 0$, the Fokker-Planck equation Eq.(11) in $r$ then can be written as:

$$\frac{\partial P(r)}{dl} = \frac{\partial}{\partial r}\left[r\frac{\partial}{\partial r}(1-r)^2 P(r)\right]P(r). \tag{12}$$

(Here we have used the same symbol $P(r)$ for the marginal probability density of $r$ as for the joint probability density $P(r(L), \theta(L))$ ). Eq.(12) has been derived earlier by several authors [10-14]. This equation also could be derived from the maximum entropy principle (MEP) [19].

The Fokker-Planck equation for a weakly disordered samples with low smaller value of reflection coefficient, that is r<<1, then Eq.(12) can be written as:

$$\frac{\partial P(r)}{\partial l} \approx \frac{\partial}{\partial r} r \frac{\partial}{\partial r} P(r), \qquad (13)$$

with the initial condition, $P(r) = \delta(r)$ for $l = 0$. The average of $r_n$ can be obtained without solving directly Eq.(13) as following. Let us define,

$$r_n = \int_0^1 P(r) r^n dr . \qquad (14)$$

Multiplying Eq.(13) by $r_n$ and integrating both sides of the equation for $r$ from 0 to 1, one gets a moment recursion equation for the average moments of the reflection coefficient,

$$\frac{dr^n}{\partial l} = n(n+1) r_n + n^2 r_{n-1}. \qquad (15)$$

Since the probability is always normalizable, the value of $r_0$ will be

$$r_0 = \int_0^1 P(r) r^0 dr = \int_0^1 P(r) dr = 1. \qquad (16)$$

Once we know the initial value $r_0$, then Eq.(16) can be solved analytically for average, square of the *rms*, and the standard deviation (*std*) of $r$ are as follows.

$$r_1(l) = l = L/\xi, \qquad (17)$$
$$r_2(l) = 2l^2, \qquad (18)$$
$$std(r(l)) = (r_2 - r_1^2)^{1/2} = l = L/\xi. \qquad (19)$$

The above expressions imply that, even for weak disordered optical media, the average reflection and STD of the reflection coefficient increase linearly with the length with same value, indicating that the reflection is not a self-averaging quantity. It is clear now why one has to consider the full probability distribution to describe better the statistical properties of the reflection coefficient of a weakly disordered media. Eq.(13) has also an analytical expression for the full distribution of $r$ for the large $l$ limit [10], which is a weak log-normal distribution.

**5. Numerical simulation details for all the disordered regimes of the FP Equation (Eq.(11))**

To get the exact form of the different probability distributions of $P(r,\theta)$, $P(r)$, and $P(\theta)$, as well as different averages of $r$ and $\theta$, one needs to solve the full Fokker-Planck (FP) equation, Eq.(11). The steps of solving the FP equation numerically are given below.

We took $r$ and $\theta$ as Cartesian variables in a two dimensional $50 \times 50$ grids (for $r$ and $\theta$). An explicit finite-difference scheme [20-21] was used to solve the FP equation. The von-Neumann stability criterion was checked and the Courant condition for the used discrete iterative length was strictly maintained. A few known results were also checked by using the rather time consuming implicit finite difference scheme.

*Allowed error bars:* Error bars of the order of $10^{-4}$ for $r$, $3 \times 10^{-3}$ for $\theta$, and $10^{-12}$ for length $l$, were allowed for the whole range of numerical calculations.

*Initial probability distribution $P(r,\theta)$ at l = 0*: The *FP* equation ( Eq.11) poses an initial value problem. The initial probability distribution *P(r, θ)* at $l = 0$ has to be specified, which will then evolve with the increase of the length of the sample. The Fokker-Planck equation Eq.(11) is however singular at *r=0*. This causes a technical problem for solving the equation numerically. To circumvent this problem, we have therefore taken an initial (fixed) scatterer with *r=.01* by putting a half-delta function potential peaked at $l = 0$ which could be physically understood as due to an initial impurity sitting at $l = 0$ ( or, the contact resistance of the leads for electronics case). By ''fixed'' we mean that it is fixed over all the realizations of the sample randomness. Phase distribution for such a weak delta-function potential will peak around $+\pi/2$ or $-\pi/2$, depending on the sign of the delta-function potential. Once $r_0 = .01$ is fixed, then the phase distribution has equal probability peak at $+\pi/2$ and $-\pi/2$. A fixed weak delta scatterer at the position $l = 0$ will not change the gross statistics, except at very smaller length scales. We have kept this initial distribution same throughout the numerical calculations. We could not consider any smaller value of the initial-fixed-reflection coefficient $(r_0)$, or a lower cut-off to the $r = 0$ singularity, for reason of convergence criterion of the numerical algorithm. An estimate can be done for the initial cut-off length $l_0$ of the sample (in terms of the localization length $\xi$) for this small $r = r_0$. Taking analytical results for the weak disordered case, one gets: $r_1(l) = r_0 = l = L/\xi$, or $l_0 = L_0/\xi \approx r_0 = .01$. This implies that the initial length is 1% of the localization length, throughout the numerical calculation. For numerical calculation, the delta-function has to be taken as the limit of a continuous function. In *(r, θ)*- space for a physically reasonable initial probability distribution, we have taken this as : $P(r, \theta)_{l=0} = \delta(r - .01) [\delta(\theta - \pi/2) + \delta(\theta + \pi/2)]$, where the delta functions are sharp Gaussians.

*Boundary conditions for r and θ for any length L ( for Eq.11):* The unit step length of the discrete evolution is taken to be $\delta l = 10^{-6}$. For every discrete evolution, we took boundary condition *P(r, θ) = 0* for *r > 1* and *r < 0* along *r* axis; and the boundary condition was taken as periodic along *θ* axis, such that $P(r, 2\pi + \delta\theta) = P(r, \delta\theta)$ for every discrete evolution.

**Initial probability distribution:** Fig.1.(a), 1.(b), and 1.(c) show this initial probability distribution *P(r, θ)* nominally at $l = 0$ *with a delta impurity*, and the marginal distribution *P(r)* of the reflection coefficient and the marginal distribution *P(θ)* of the phase *θ*. It should be emphasized again here that the initial probability distribution of the phase, *θ,* can be taken at any small enough length. However, the statistical properties of the system do not depend on the initial distribution except for initial very small lengths.

## 6. Results and discussions

### *6. A. Evolution of P(r, θ) with the length L for different disorder strengths*

We will consider the evolution of the full probability distribution *P(r, θ)* for different dimensionless lengths *l (=L/ξ)* for the three main regimes of disorder strength : i) weak *(2kξ>> 1)*, ii) intermediate/medium *(2kξ~ 1)*, and iii) strong *(2kξ~ 1)*, defined by the disordered strength parameter *2kξ*.

*A.1. P(r, θ) for weak disorder : 2kξ = 100 (2kξ>> 1):* Figs.2(a), 2(b), 2(c), and 2(d) show typical distributions of the joint probability in *(r, θ)*-space for different lengths *l = 1, 2, 5,* and *10*. Evolution shows clearly that the phase part is uniformly distributed. Evolution of the probability is mainly along the amplitude *(r)* axis. Probability evolves with length and peaks near *r = 1* for large lengths. These evolution pictures imply that the random (uniform) phase distribution holds for the weak disorder very well, or for the large values of the localization length. Later we show the nearly exact range for the disorder parameter *(2kξ)* where the random phase approximation is valid, for a given length of the sample.

*A.2. P(r, θ) for intermediate/medium disorder: 2kξ = 1 (2kξ= 1):* Similarly, Figs. 3(a), 3(b), 3(c), and 3(d) are as Fig.2, but for the disorder strength parameter *2kξ = 1*, i.e., medium/intermediate disorder regime. These evolutions clearly show that the probability distribution *P(r, θ)* is quite complex and has an asymmetric distribution in phase. The phase distribution peaks on one side *(2π<θ <π)* for large lengths.

*A.3. P(r, θ) for strong disorder : 2kξ = .001 (2kξ << 1) :* Similarly, Figs. 4(a), 4(b), 4(c), and 4(d) are as Fig.2, but for the disorder parameter 2kξ = .001, i.e., the strong disorder regime. Probability distribution is perfectly symmetric centered at the phase $\theta = \pi$. In the strong disorder parameter regime the sample tries to behave as a nearly a *perfect reflector* and totally reflect back the incoming wave with opposite phase. Evolution pictures show that *P(r, θ)* peaks around *r = 1* and is symmetrically around $\theta = \pi$ for larger lengths. It will be shown later that the distribution does not change substantially with further increase of the disorder strength parameter, i.e. the distribution is insensitive to the disorder parameter $2k\xi$ in that regime.

*6.B. Marginal distribution P(r) of the reflection coefficient (r) and Marginal distribution P(θ) of the phase (θ) for: (a) weak, (b) medium/intermediate, and (c) strong disorder regimes*

In Fig. 5, (i) subplots Fig.5(a), 5(b), and 5(c) show the marginal probability distribution $P(r)$ when the phase (θ) part of the $P(r, \theta)$ is integrated out, and (ii) subplots 5(a'),5(b'), and 5(c') corresponding marginal distributions $P(\theta)$ of the phase θ when r part of the full distribution $P(r, \theta)$ is integrated out for the three cases of disorder regimes considered above are the (a) weak, (b) medium/intermediate, and (c) strong. Figs. 5(a')-(c') clearly indicate that the phase *(θ)* distributions $P(\theta)s$ are very different for the three different regimes of disorder parameters considered here. However, as shown in Figs. 5(a) & (b), the reflection coefficient *(r)* distribution does show reasonable variation with the strength of the disorder parameter. Although, they are quite different for smaller length (*l*~1), however they are quite similar for larger lengths (*l*>>1). The behavior of the probability distribution indicates that the distribution of the phase has some effect on the distribution of the reflection coefficient $P(r)$, depending on the length and disorder strength, hence on the average transport properties.

*6.C. Distribution of the phase (θ) with the strength of disorder parameter 2kξ*

We now study the phase distribution $P(\theta)$ with the strength of the disorder parameter *(2kξ)*. We consider here mainly two cases of fixed lengths: *1) l = 1*, and 2) the asymptotic limit when $l \rightarrow \infty$.

*C.1. The phase (θ) distribution P(θ) for different disorder strength parameters (2kξ) for fixed length l = 1: RPA regime*

Fig.6 shows the plot of $P(\theta)$ for the fixed length *l = 1*, with different values of the disorder strength parameters *(2kξ)*. Figures show that in the strong as well as in the weak disorder limit, $P(\theta)$ is insensitive to the strength of the disorder parameter *2kξ*, when the length is *l*~1 or more. In particular, the phase distribution is uniform for the weak disorder case, and it has a perfectly symmetric peak centered at $\theta = \pi$ for the strong disorder case. In the case of intermediate disorder, the distributions are asymmetric with respect to the $\theta = \pi$ point. For the medium/intermediated disorder strength parameter, $P(\theta)$ interpolates continuously between the strong and the weak disorder limiting phase distributions.

**RPA validation range:** It is clear from the $P(\theta)$ distributions, that the random phase approximation (RPA) is valid for the condition $\xi > \lambda = 2\pi/k$, where λ is the wave length of the incoming wave and ξ is the localization length. *This has been checked systematically in our numerical calculations.*

*C.2. The phase (θ) distribution P(θ) in the asymptotic limit of large lengths l→α:*

Solution of the FP equation ( Eq.(12) ) gives the full probability distribution $P(r, \theta)$ in the *(r, θ)-space*. For larger lengths, one can make the approximation: $r \approx 1$ and $P(r) \approx \delta(r - 1)$. In this asymptotic limit, marginal distribution for phase $P(\theta)$ becomes:

$$P(\theta) = \int_0^1 P(r,\theta)\delta(r-1)dr \equiv P(1,\theta). \tag{20}$$

Now from Eq.(11) and Eq.(20), we get the marginal probability distribution of phase for larger length limit as:

$$\frac{\partial P(\theta)}{dl} = \frac{\partial}{\partial \theta}\left[(1+\cos\theta)\frac{\partial}{\partial \theta}(1+\cos\theta)\right]P(\theta) - 2k\xi\frac{\partial P(\theta)}{\partial \theta} \quad . \tag{21}$$

Fig.7 shows the plot of $P(\theta)$ in the asymptotic limit for different disorder strength parameters *(2kξ)*, as the parameter varies from the weak to the strong disorder limits.

Plots in Fig. 7, for large length (i.e., r≈1), show that in the interval from *0* to *2π*, (i) a uniform phase distribution in the weak disorder regime; (ii) an asymmetric phase distribution in the intermediate disorder regime; and (iii) a symmetric double peaked distribution around $\pi$ in the strong disorder regime. These distributions indicate that the essential features of $P(\theta)$ have not changed with the $r \approx 1$ approximation, relative to the *l = 1* case which is shown in Fig.6.

## 7. Review of the previously reported results on phase distribution and comparison with the present work

We show in our numerical simulations that the random phase approximation (RPA) is valid for localization length larger than the wavelength, i.e. $\xi > \lambda$. Physical meaning of the RPA with this condition is that the incoming wave has to undergo multiple sub-reflections before escaping a localization length and in the process the wave randomizes its phase. This weak disorder regime also can be expressed for a short sample length or within the Born single scattering approximation or regime. For the weak disorder case, the localization length is larger than the wavelength. Hence the phase of the wave gets randomized. In the other extreme case of the strong disorder, system tries to behave as a perfect reflector. Hence, phase of the reflected wave tries to peak at $\pi$ (opposite phase with respect to the incoming wave). In the regime of intermediate disorder, $P(\theta)$ distribution is disorder specific and has some bias points to peak at $\theta > \pi$. The value of the peaking point of the $\theta$ distribution at $\theta > \pi$ can be interpreted as that mostly uniform distributions, however, waves are less back reflected, rather penetrating deep before the reflection as the distribution is peaked above $\theta > \pi$.

At this point we will discuss the results that are previously reported in the literature. Phase distribution for *1D* systems have been studied earlier by Sulem [22], Stone *et al.* [23], Jayannavar [24], Heinrichs [25], and Manna *et al.* [26]. Sulem's result missed the random phase distribution in the limit of weak disorder, however, the calculated phase distributions show peak near $\pm \pi$, and it does not show a symmetric distribution like the present work. The study of Stone *et al.* showed for the 1D Anderson model that (i) for the case of weak disorder the phase distribution is uniform, (ii) non-uniform for the strong disorder, (iii) and pinning of phase distribution near $2\pi$ for very strong disorder and large lengths. Their work also confirm that the distribution of the phase is insensitive to the disorder strength in the limit of weak as well as strong disorder limits. Manna *et al.* show uniform phase distribution from *0* to *2π* for the weak disorder, peaking of the phase distribution near $\pi$ for the strong disorder, agreeing with the present work. Jayannavar's calculation for the asymptotic large length (i.e., $l \to \infty$) phase distribution shows a uniform distribution of phase for weak disorder. However, the phase distribution for strong disorder shows several peaks that does not agree with our results.

## 8. *Marginal probability distribution, P(r), of the reflection coefficient, r, with respect to the disorder strength parameter 2kξ*

Fig.8 shows the probability distribution, *P(r)*, of the reflection coefficient, *r*, with the disorder parameter strength *(2kξ)* for the sample length *l = 1*. The distributions show weak dependence on the strength of the disorder parameter. But the distribution certainly has a small spread, not drastic. *It can be seen that though the phase (θ) distributions P(θ) is quite different for different strengths of the disorder parameter (2kξ), the reflection coefficient (r) distribution P(r) does not change appreciably with the disorder strength for a fixed length of the sample. However, there is a finite and smaller effects.*

## 9. Conclusions and Remarks
### 9.1 Summary of the results

We have solved here the *1D* optical transport problem for the case of Gaussian white-noise optical disorder potential numerically, by solving joint probability distributions. This is almost a complete solution of transport properties of the *1D* Gaussian white-noise random potential that goes beyond the conventional random phase approximation (RPA), or uniform phase distribution, valid only for weak disorder. We have evolved the full probability distribution in the reflection coefficient (*r*) and the associated phase (*θ*) space (i.e. *(r, θ)-space*) of the complex reflection amplitude $R = \sqrt{r} e^{i\theta}$ for a 1D disordered sample, with different lengths and with different disorder strengths. For our numerical solution, we have taken a fixed initial reflection coefficient $r_0 = .01$ for all realization of disorder as the Fokker-Planck equation for $P(r, \theta)$ is singular at $r = 0$. Gross statistical properties of the system are not expected to change with this weak extra scatterer. It may, however, affect the very sensitive details corresponding to the limit $r \to 0$ at $l \to 0$. Our numerical work is a systematic study to observe the contribution of the phase fluctuations to reflection coefficient averages.

On the basis of the results obtained, our conclusions are the following:

**(A)** *Phase θ distribution P(θ) in different regimes of disorder*

1). The *random phase approximation* (RPA) implying uniform phase distribution over $2\pi$ is valid for the condition $\xi/\lambda > 1$ ($\xi$ is the localization length and $\lambda$ is incoming wavelength), that is, in the weak disorder limit. Physically, this means that the wave has to undergo multiple sub-reflections before it moves through one localization length or scattering mean free path. $P(\theta)$ is independent of the disorder in the weak disorder limit ($2k\xi >> 1$).

2). In the intermediate disorder regime, phase distribution is complex, and asymmetric about $\theta = \pi$ point in the interval 0 and $2\pi$, peaking in between $\pi$ and $2\pi$. Also, the phase distributions are strongly disorder strength parameter, $2k\xi$, dependent. In this regime back reflection is less prominent than the uniform distribution, however, the wave get back reflected after a deep penetration. This provides us understanding of the peak of the reflected wave around $2\pi$.

3). In the strong disorder regime the distribution of the phase is perfectly symmetric in the interval from 0 and $2\pi$, centering at $\theta = \pi$. This arises due to the opposite phase of the back reflected wave in the disordered regime. The distribution is independent of the disorder in the strong disorder limit ($2k\xi << 1$) even for smaller sample lengths.

**(B) Reflection coefficient (*r*) distributions *P(r)***

The probability distribution for the reflection coefficient peaks at $r = 1$ for large lengths $l \equiv L/\xi >> 1$. Though probability distribution $P(\theta)$ for the phase varies qualitatively with the variation of the disorder strength, the probability distribution for the reflection coefficient $P(r)$ does not change qualitatively with the disorder strength for a fixed length of the sample. In particular phase distribution has a weaker effects on the reflection coefficient distribution.

*9.2 Extension of the phase distribution to electronic case*

For 1D, the Schrödinger and the Maxwell equations projects to the Helmholtz equation, and they are similar to each other, only the form of the potentials are different. The Landauer four-probe resistance formula [27] shows that the dimensionless resistance as a function of the reflection coefficient:

$$\rho(L) = r(L)/(1 - r(L))$$ .

Making the transformation from *P(r)* to *P(ρ)* by using *P(ρ) = P(r)(dr/dρ)*, one obtains the Fokker-Planck equation for the resistance, and the mean and the *STD* of the resistance can be derived from the above variable change as shown in details in [12]. The functional change in the magnitude of the reflection will not change the form of the value of the phase distribution, therefore, the phase distribution of the rasistance will be same as the phase distribution of the reflection coefficient.

*9.3 Applications of the phase statistics*

1D disordered media is equivalent to the continuum of the discrete random layered media. These types of media appear in optical media in semiconductors, layered electronic media, atmospheric dielectric layered media for different electromagnetic wave transportation through our atmosphere, underground dielectric layered media in search of earth's underground oils. The reflection is associate with phase, therefore, it is important to know the actual phase distribution, to better predict the regime of the reflection. 1D disordered media may be also used for the phase randomizer for electrical or electronics cases. Recently 1D disordered media is used for one-dimensional multi-channel analysis of the light reflected from biological cells for ultra-early cancer detection [28,29], as well as well as transport in stochastic absorbing media [14]. Knowing 1D phase distribution would be helpful for accurate parameterization of the reflection coefficients as well as understanding of the random layered media.

**Declaration:**

The author(s) declare(s) that there is no conflict of interest regarding the publication of this article.

**Acknowledgements**

The author gratefully acknowledge N. Kumar for stimulating discussions. The work was partially supported by the NIH grant No. R01EB003682. The author also acknowledge financial support from the University of Memphis.

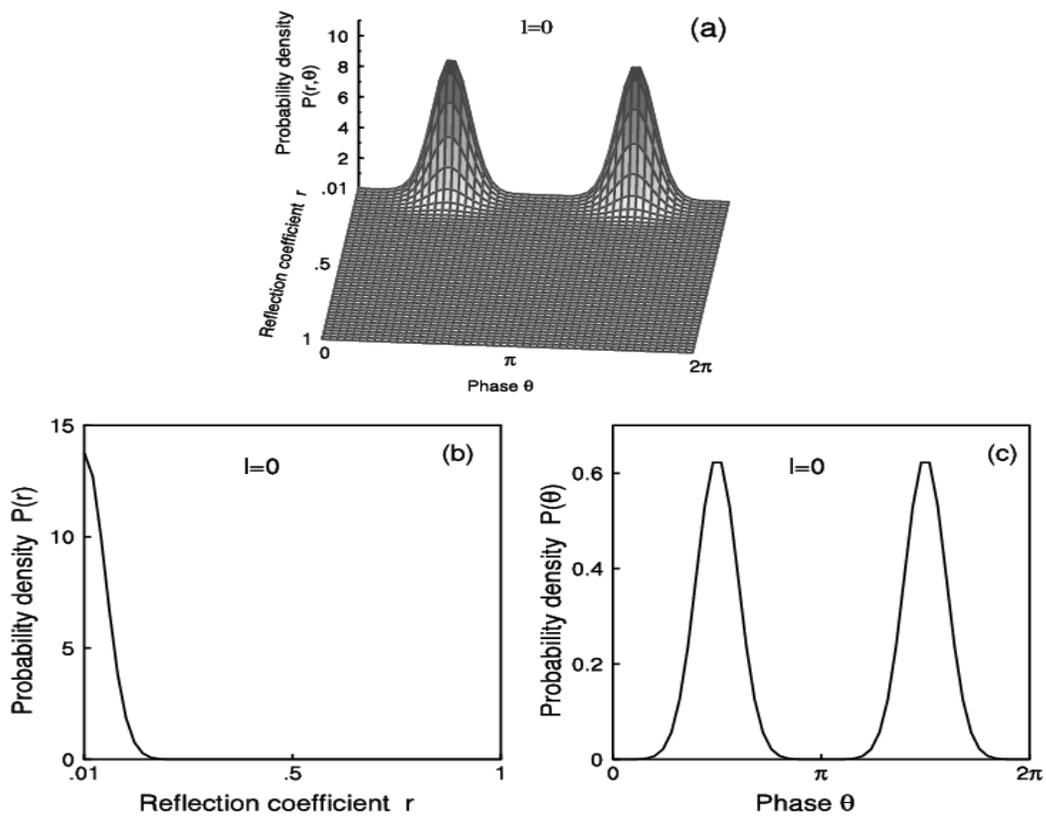

Fig. 1. Initial probability distribution, i.e., when the sample length $l = 0$: (a) Probability distribution $P(r,\theta)$. (b) Marginal reflection probability distribution $P(r)$. (c) Marginal phase probability distribution $P(\theta)$.

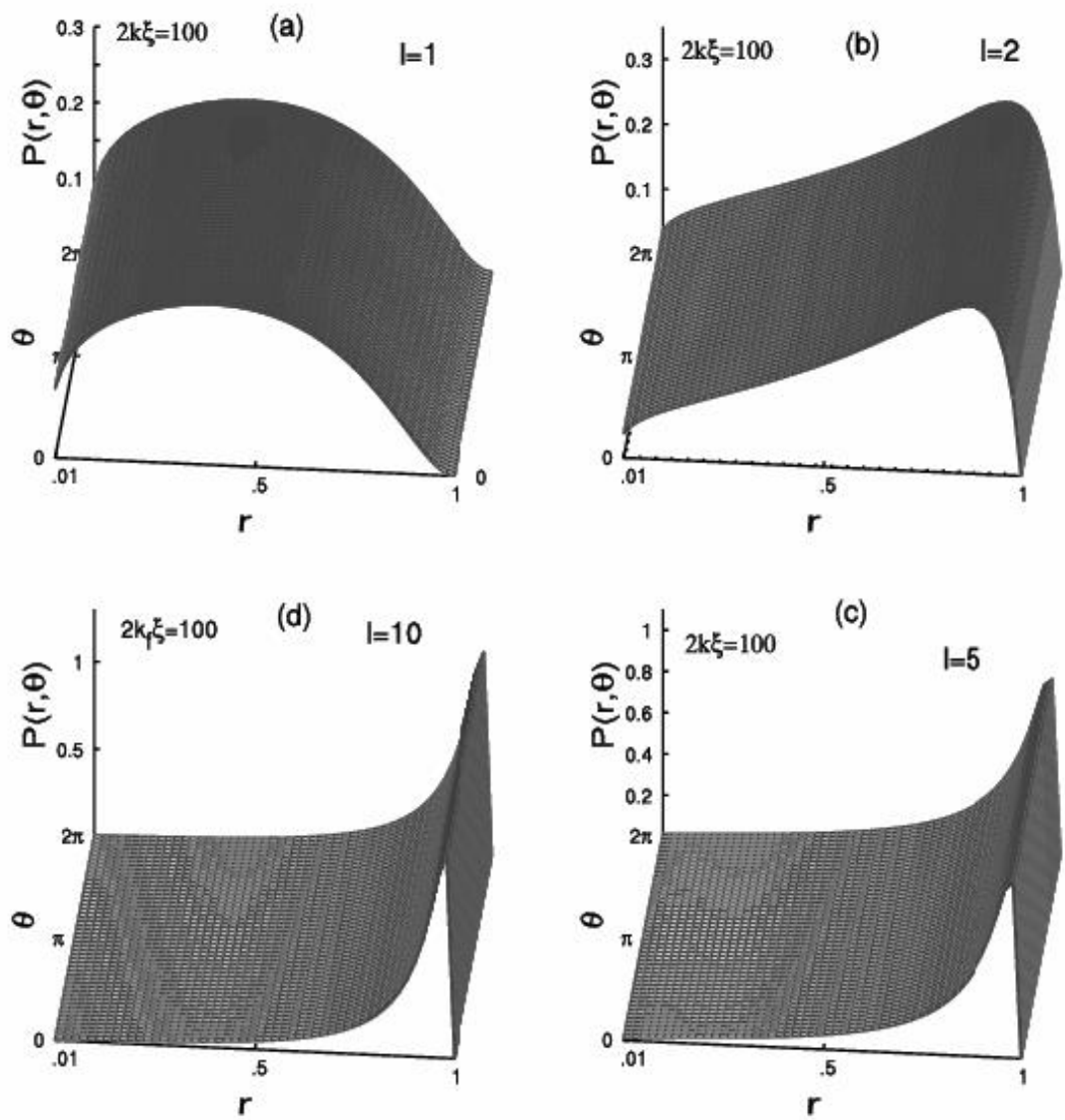

Fig. 2. Evolution of the probability $P(r,\theta)$ with the sample length $l\ (=L/\xi)$ in the weak disorder regime for a fixed disorder strength parameter $2k\xi = 100$. Plots are for sample lengths: (a) $l = 1$, (b) $l = 2$, (c) $l = 5$, and (d) $l = 10$.

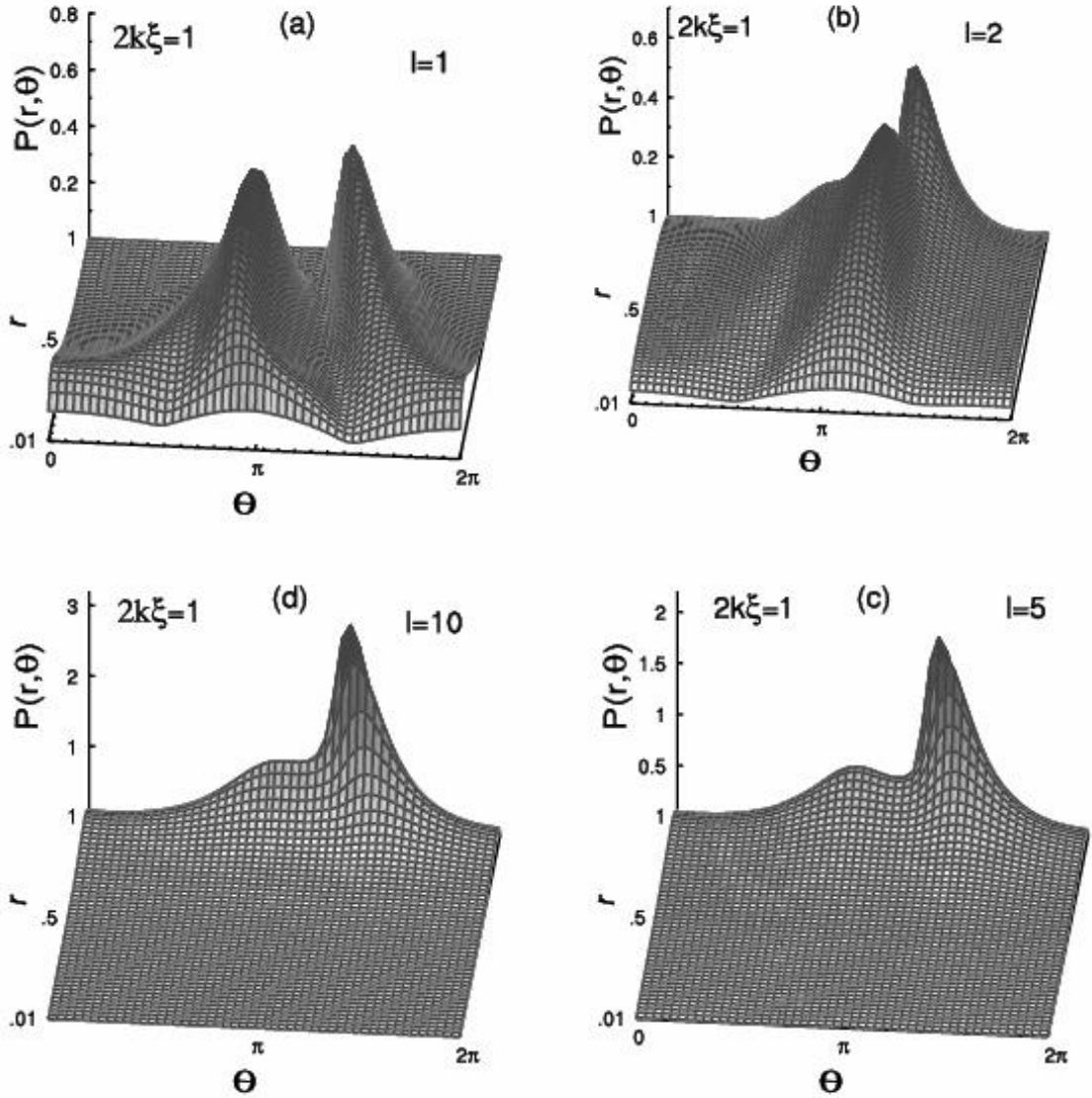

Fig.3. Evolution of the probability $P(r,\theta)$ with the sample length $l$ $(=L/\xi)$ in the medium/intermediate disorder regime for a fixed disorder strength parameter $2k\xi= 1$. Plots are for sample lengths: (a) $l = 1$, (b) $l = 2$, (c) $l = 5$, and (d) $l = 10$.

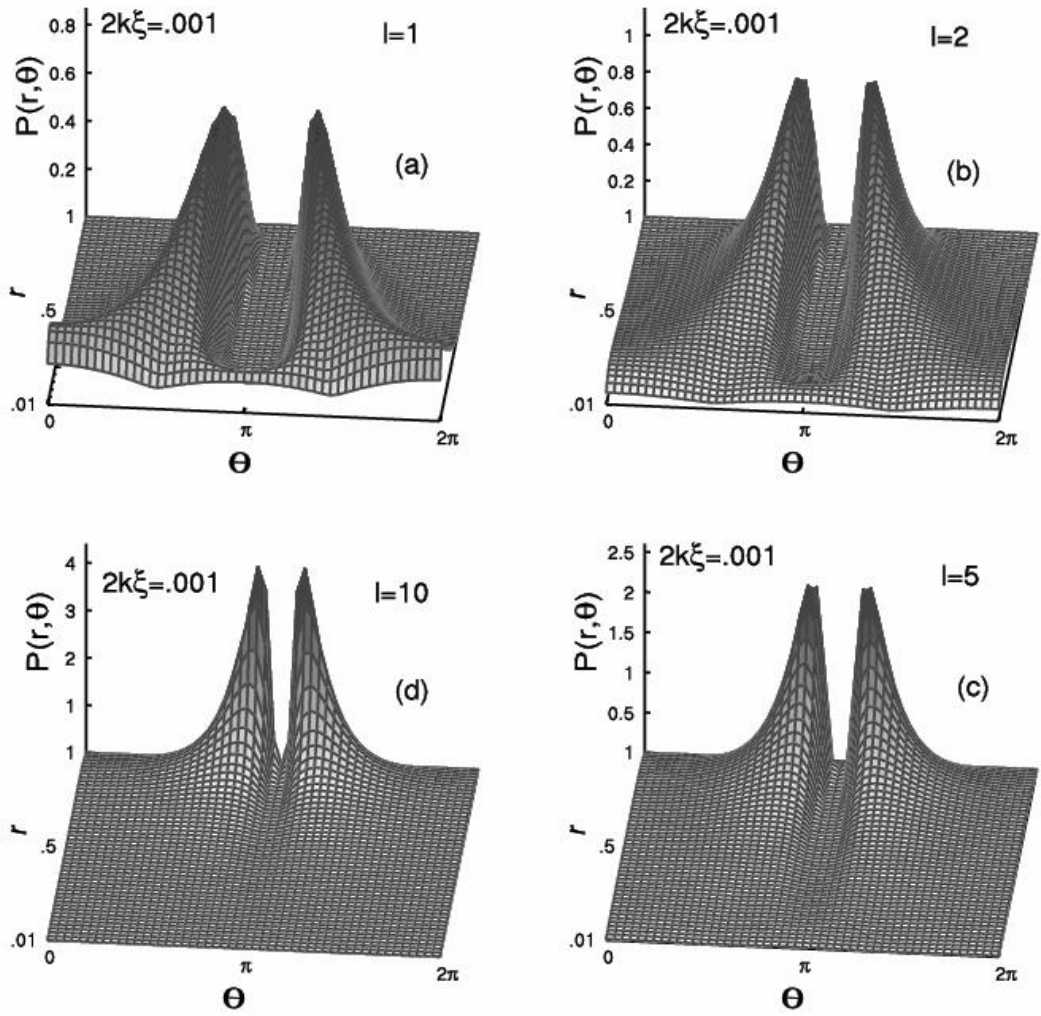

Fig. 4. Evolution of the probability $P(r, \theta)$ with the sample length l $(=L/\xi)$ in the strong disorder regime for a fixed disorder strength parameter $2k\xi = .001$. Plots are for sample lengths (a) $l = 1$, (b) $l = 2$, (c) $l = 5$, and (d) $l = 10$.

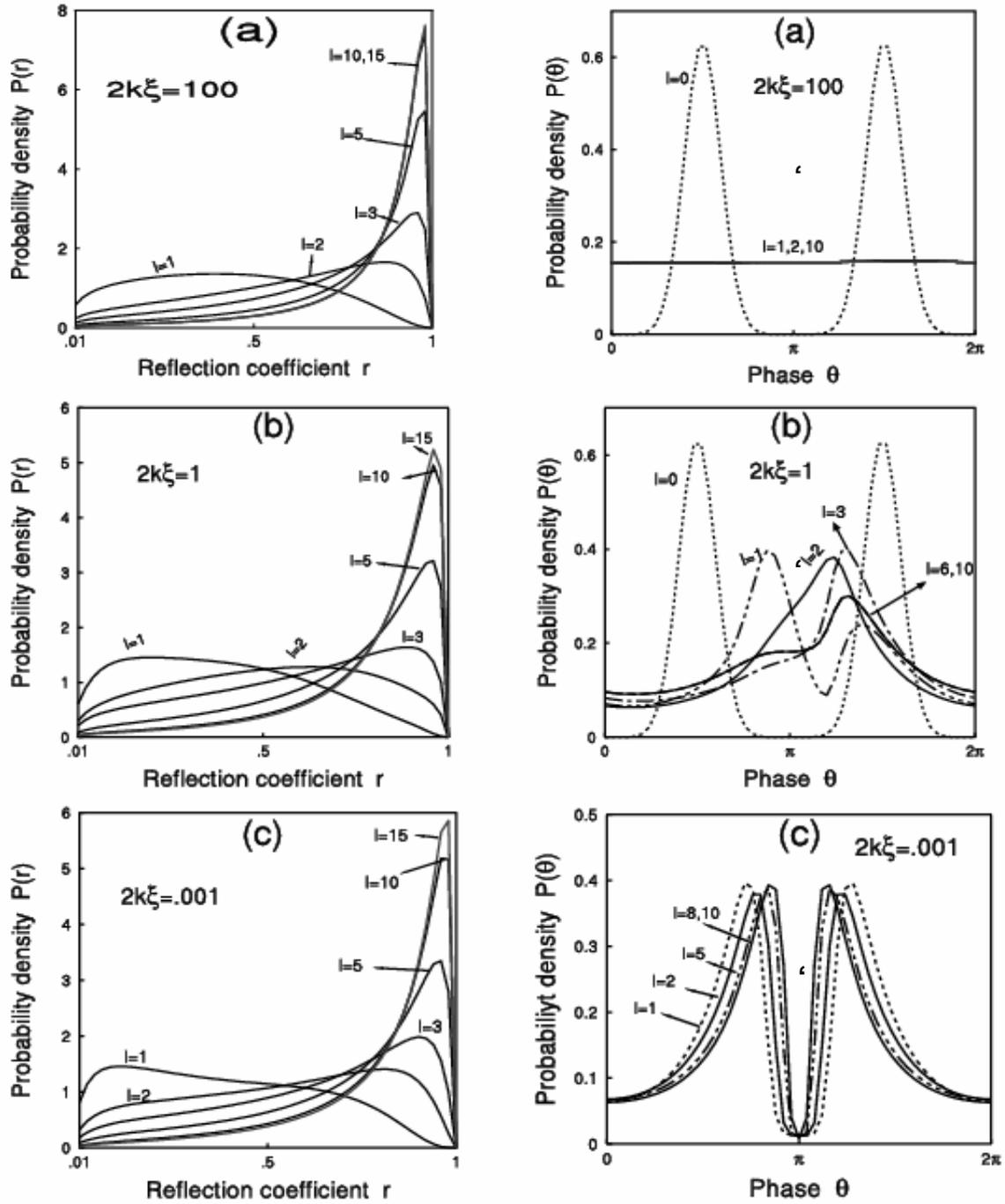

Fig. 5. Marginal probability distribution *P(r)* for the fixed disorder strength parameters: (a) Weak disorder, $2k\xi = 100$, (b) Intermediate/medium disorder, $2k\xi = 1$, and (c) Strong disorder, $2k\xi = .001$ (corresponding to Figs. 2, 3 and 4) for different sample lengths. Corresponding marginal phase distributions *P(θ)* for : (a') Weak disorder, $2k\xi = 100$, (b') Intermediate/medium disorder, $2k\xi = 1$, and (c') Strong disorder, $2k\xi = .001$ for the different sample lengths *l*.

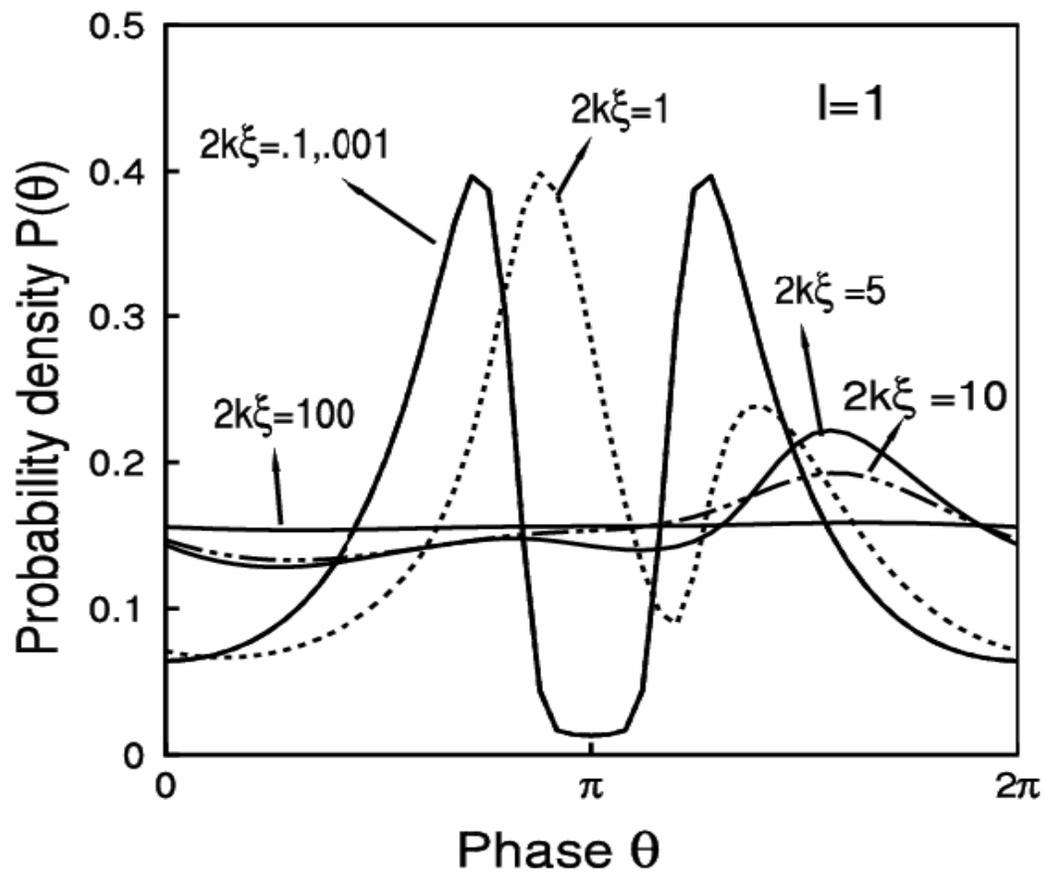

Fig. 6. Phase θ distribution $P(\theta)$ against the disorder parameter $2k\xi$ for a fixed sample length $l = 1$.

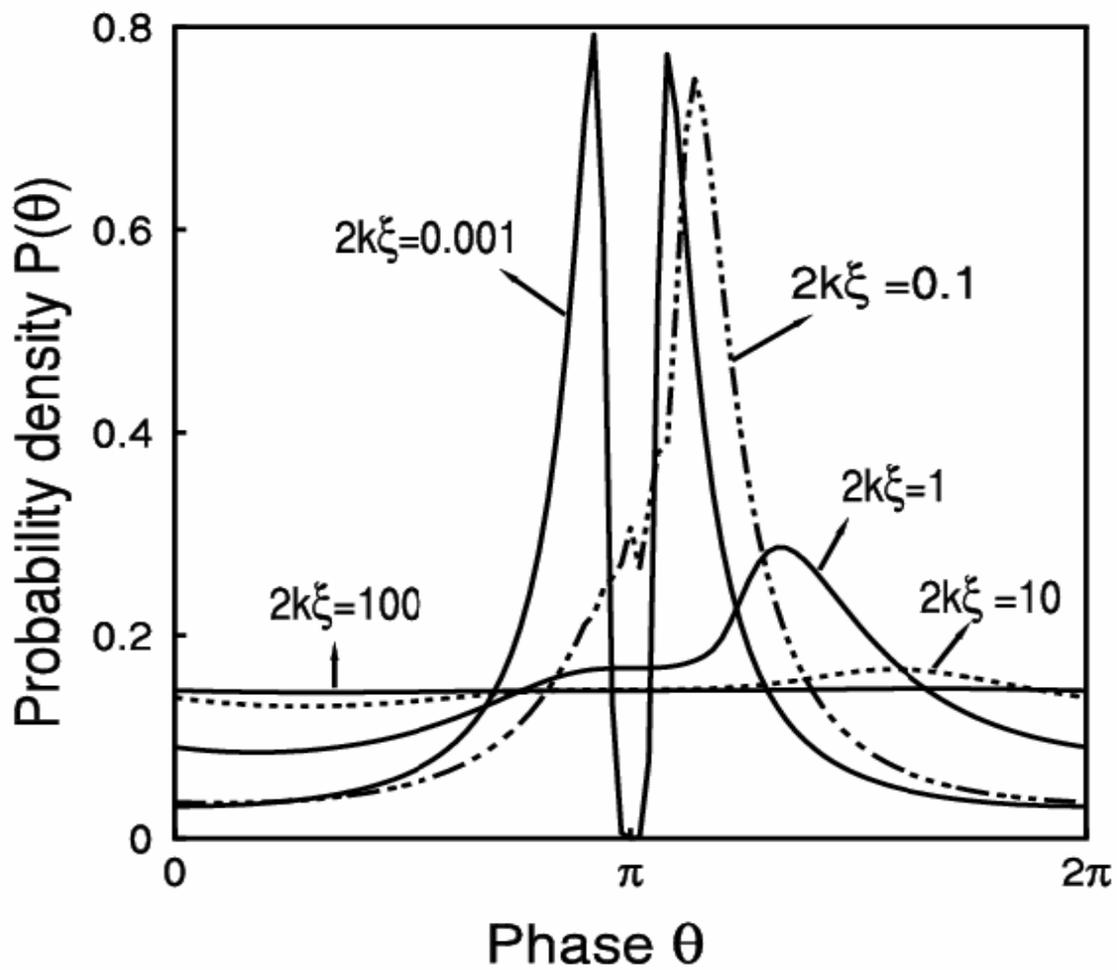

Fig. 7. Phase θ distribution *P(θ)* against the disorder parameter *2kξ* for the asymptotic limit of large length *l* = 10 (large).

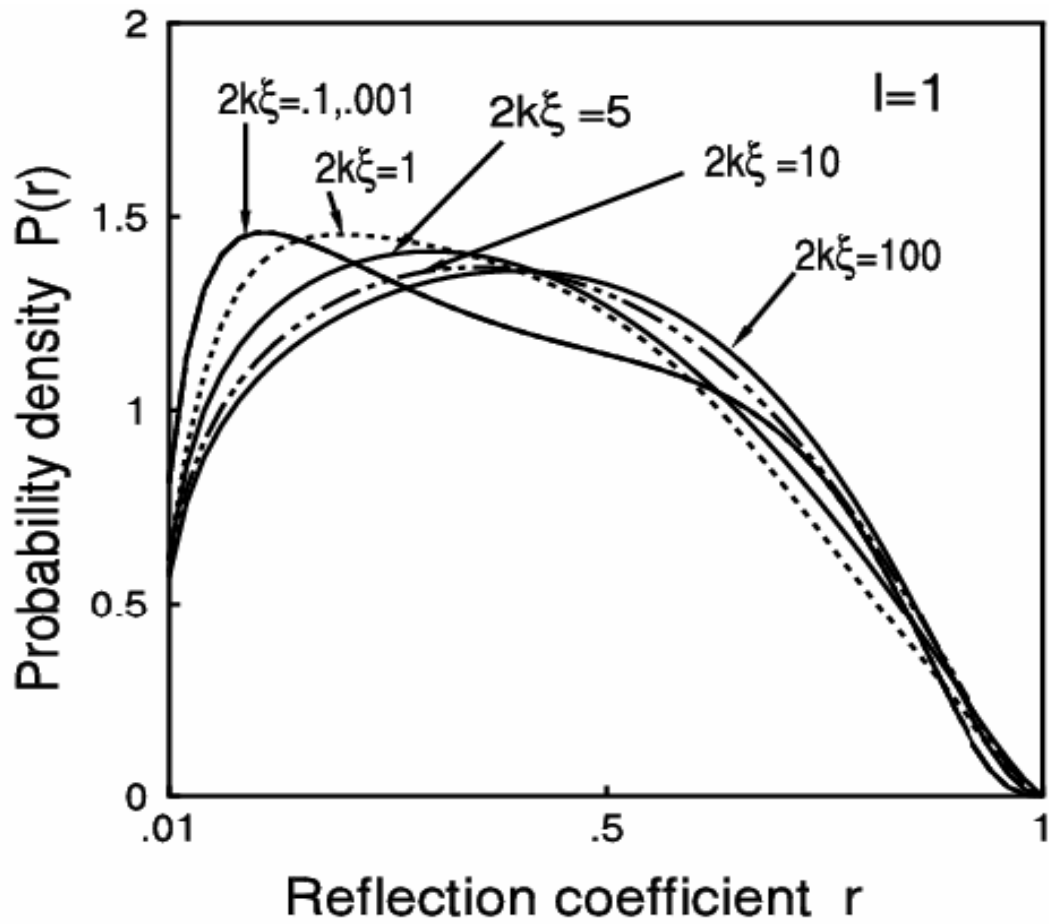

Fig. 8. Reflection coefficient *r* distribution *P(r)* against the disorder strenght parameter *2k$\xi$* for a fixed sample length *l* = 1.